\documentclass[aps,pra,twocolumn,showpacs,showkeys,superscriptaddress,nobalancelastpage,nofootinbib,floatfix,longbibliography]{revtex4-2}
\usepackage{graphicx}
\usepackage{color}
\usepackage{dcolumn}% Align table columns on decimal point
\usepackage{latexsym}
\usepackage{cancel}
\usepackage[normalem]{ulem}
\usepackage{hyperref,amssymb}
\usepackage{url}
\usepackage{color,soul}
\usepackage{graphicx}
\usepackage{verbatim}
\usepackage{multirow}
\usepackage{amsmath}
\newcommand{\beq}{\begin{eqnarray}}
\newcommand{\eeq}{\end{eqnarray}}
\usepackage{mathrsfs}
\usepackage{float,soul}
\usepackage[dvipsnames]{xcolor}
\usepackage{mathtools}
\usepackage{slashed}
\usepackage{graphicx}   % need for Fig.ures
\usepackage{epstopdf}
\usepackage{subfig} % use for side-by-side Fig.ures
\usepackage{hyperref}   % use for hypertext links, including those to external documents and URLs
\usepackage{bbold}
\usepackage{wasysym}
\usepackage{feynmp}
\usepackage{tikz}
\usepackage{times}
\begin{document}
\title{Parity-Time Symmetric Spin-1/2 Richardson-Gaudin Models }

\author{M. W. AlMasri} 
\email[]{mwalmasri2003@gmail.com}
\affiliation{Wilczek Quantum Center, School of Physics and Astronomy, Shanghai Jiao Tong University, Shanghai 200240, China}

\date{\today}

\begin{abstract}
We construct a  $\mathcal{PT}$-symmetric Richardson--Gaudin models for spin-$\tfrac{1}{2}$ systems by deforming the closed integrable Hamiltonian through complex-valued transverse magnetic fields and coupling constants. By defining parity as $\mathcal{P} = \prod_i \sigma_i^z$ and adopting a time-reversal operator that flips only the $y$-component of spin, we establish a consistent $\mathcal{PT}$-symmetric framework distinct from open-system approaches based on Lindblad dynamics. The resulting model remains integrable, with conserved charges satisfying generalized commutativity conditions. We explicitly construct the Hermitian counterpart via a similarity transformation and identify the metric operator $\rho = e^{-\sum_i q_i S_i^z}$ that defines the physical inner product. Numerical diagonalization reveals the characteristic $\mathcal{PT}$ spectral structure: eigenvalues are either real or form complex conjugate pairs, with partial symmetry breaking wherein low-energy states remain in the unbroken phase. We further derive exact analytical expressions for spin dynamics, showing coherent oscillations in the unbroken phase and exponentially modulated behavior in the broken phase. 
\end{abstract}
\

\maketitle

\section{Introduction}
The study of many-body quantum systems has been a cornerstone of physics, providing deep insights into the behavior of interacting particles in condensed matter, nuclear physics, and quantum information theory \cite{many,nuclear,fazio}. Among the diverse array of models developed to describe such systems, the Richardson-Gaudin (RG) model stands out as a remarkable example of an exactly solvable system \cite{RMP,thesis}. Originating from the pioneering work of Richardson in the context of superconductivity and later extended by Gaudin to encompass a broader class of integrable systems, the RG model provides an interesting framework for understanding pairing interactions and collective phenomena in quantum systems \cite{R1,R2,R3,R4,Gaudin, Bethe}. Sklyanin's work introduces the separation of variables technique in the Gaudin model, providing a foundation for solving spectral problems in integrable systems \cite{Sklyanin}. Feigin, Frenkel, and Reshetikhin extend this framework by connecting the Gaudin model to the Bethe Ansatz and exploring its behavior at the critical level \cite{Frenkel}. Cambiaggio et al. investigate the integrability of pairing Hamiltonians, which are closely related to Richardson-Gaudin models, while more recent works by Skrypnyk generalize these ideas to elliptic and quasi-trigonometric r-matrices, broadening their applicability \cite{Cambiaggio,Skrypnyk, Skrypnyk1}. Dimo and Faribault, along with collaborators, delve into quadratic operator relations and Bethe equations, advancing the understanding of spin-1/2 Richardson-Gaudin models under anisotropic conditions and external magnetic fields \cite{Dimo, Faribault} . In \cite{Villazon}, the authors demonstrated that the fully anisotropic central spin model with XX Heisenberg interactions is integrable, revealing an extensive set of conserved quantities and exact Bethe ansatz eigenstates that split into bright and dark states, the latter enabling robust quantum memory and limiting spin bath polarization. In \cite{Claeys}, the authors  solve an open quantum system with collective dissipation by mapping its Liouvillian to a non-Hermitian XXZ Richardson-Gaudin model, revealing a pseudo-Hermitian $\mathcal{PT}$-symmetric spectrum that undergoes dissipative quantum phase transitions through exceptional points, supports a nontrivial steady state, and exhibits quantized eigenvalue ratios and logarithmic spectral gap scaling in the strong-coupling regime. In \cite{Nadai}, the integrability of the spin-1 $XX$ central spin model in a tilted magnetic field was proved by constructing its complete set of conserved charges and their polynomial relations, and the emergence of dark states at strong coupling was numerically demonstrated, analogous to the spin-1/2 case. 

\vskip 5mm

Open quantum systems describe physical systems that interact with an external environment. In contrast to closed systems, which evolve unitarily under Hermitian Hamiltonians, open systems exhibit non-unitary dynamics due to energy exchange, dissipation, and decoherence induced by their coupling to the environment. This interaction leads to irreversible processes such as population decay, loss of quantum coherence, and measurement-induced backaction—phenomena that cannot be captured within the framework of standard Hermitian quantum mechanics.

A common approach to modeling such systems is through an effective non-Hermitian Hamiltonian \cite{Rotter}:
\begin{equation}
H_{\text{eff}} = H_0 - i\Gamma,
\end{equation}
where $H_0$ is a Hermitian operator governing the coherent dynamics of the system, and $\Gamma \geq 0$ is a positive semi-definite operator encoding dissipative and decohering effects. The imaginary term $-i\Gamma$ results in a non-unitary time evolution, manifesting physically as a decay in the norm of the state vector—representing the loss of probability amplitude from the system into the environment. While the effective non-Hermitian description provides valuable insight into dissipative dynamics, a more rigorous treatment of open quantum systems employs the density matrix formalism. The time evolution of the density operator $\rho$ is governed by a quantum master equation. For Markovian environments—where memory effects are negligible—the dynamics are described by the Lindblad master equation \cite{davies,breuer,rivas,rosario}:
\begin{equation}
\frac{\partial \rho}{\partial t} = -\frac{i}{\hbar} [H, \rho] + \sum_k \left( L_k \rho L_k^\dagger - \frac{1}{2} \{L_k^\dagger L_k, \rho\} \right),
\end{equation}
where $H$ is the Hermitian system Hamiltonian, $L_k$ are the Lindblad (or jump) operators representing environmental interactions, and $\{A, B\} = AB + BA$ denotes the anticommutator. The Lindblad form guarantees complete positivity and trace preservation of $\rho(t)$, ensuring a physically consistent description of the system's evolution. Under certain approximations—such as the weak-coupling limit, adiabatic elimination of fast-decaying modes, or proximity to a steady state—the full Lindblad dynamics can be reduced to an effective non-Hermitian Hamiltonian description. For instance, in a two-level system coupled to a thermal bath, spontaneous emission and absorption processes are modeled by Lindblad operators $L_\downarrow$ and $L_\uparrow$. In the regime of weak dissipation, the reduced dynamics may be approximated by:
\begin{equation}
H_{\text{eff}} = H_0 - i\gamma,
\end{equation}
with $\gamma > 0$ representing the net decay rate. This illustrates how non-Hermitian physics naturally emerges from the more general framework of open quantum systems.
\vskip 5mm
It is within this broader context that $\mathcal{PT}$-symmetric quantum mechanics finds its physical grounding. $\mathcal{PT}$ symmetry—where a non-Hermitian Hamiltonian is invariant under the combined action of parity ($\mathcal{P}$) and time-reversal ($\mathcal{T}$) symmetries—represents a special class of open quantum systems in which gain and loss are balanced in a spatially symmetric fashion. While generic open systems break both Hermiticity and symmetry, $\mathcal{PT}$-symmetric models occupy a distinguished subspace where the competition between coherent dynamics and structured dissipation leads to entirely real energy spectra, provided the symmetry is unbroken \cite{Bender,Bender1,Dorey,Mostafazadeh,Jones, review}. A Hamiltonian $H$ is $\mathcal{PT}$-symmetric if it commutes with the $\mathcal{PT}$-operator, i.e. $[H,\mathcal{PT}]=0$. The main applications of $\mathcal{PT}$-symmetric quantum mechanics are found in optics \cite{Guo}, topological phases of matter \cite{Esaki}, Bose-Einstein condensation \cite{BEC}, spin chains \cite{Gehlen, Korff, Fring, Vinokur, Prosen, Eremin, Wahiddin}, integrable systems \cite{Fring1}, $\mathcal{PT}$-symmetric qubits and anti-$\mathcal{PT}$ qubits \cite{qubit,anti}, $\mathcal{PT}$-symmetric Rabi model for qubit interacting with classical light \cite{Rabi} and quantum light \cite{Rabi1},  interatomic interaction models such as the $\mathcal{PT}$-symmetric Morse potential \cite{Almasri}, $\mathcal{PT}$-symmetric Cooper pairing in superconductors \cite{Das}, and many other areas. 

    \vskip 5mm

In a seminal work, Babelon and Talalaev \cite{Babelon} demonstrated that the spectral problem of certain Jaynes–Cummings–Gaudin (JCG) models can be solved directly in terms of the eigenvalues of conserved charges, yielding quadratic Bethe equations without the need to determine individual Bethe roots—a step that circumvents a major bottleneck in traditional Bethe ansatz methods.

This charge-based method not only simplifies spectral analysis but also reveals deeper algebraic structures underlying integrability. However, its application has so far been limited to Hermitian systems. In parallel, non-Hermitian quantum mechanics—particularly $\mathcal{PT}$-symmetric models—has emerged as a powerful framework for describing open quantum systems, exceptional points, and novel phases of matter with real energy spectra despite non-Hermiticity. Yet, the interplay between integrability and $\mathcal{PT}$ symmetry remains largely unexplored in the context of RG models especially for spin-1/2 case in arbitrary magnetic field. 

Motivated by these developments, we extend the conserved charges approach to $\mathcal{PT}$-symmetric Richardson–Gaudin models. This extension is nontrivial: the presence of complex couplings and non-Hermitian interactions modifies the algebraic structure of the conserved charges and challenges the reality of the spectrum. By adapting the Babelon–Talalaev framework to this setting, we aim to establish conditions under which $\mathcal{PT}$-symmetric RG models remain integrable,  and uncover how $\mathcal{PT}$ symmetry influences the distribution of eigenvalues. Our work thus bridges two frontiers—exact solvability and non-Hermitian physics—opening new pathways for understanding integrable dynamics in open and engineered quantum systems.
   \vskip 5mm

The organization of this paper is as follows. 
In Section II, we describe the Richardson-Gaudin models. 
In Section III, we introduce the $\mathcal{PT}$-symmetric Richardson-Gaudin model and compute the eigenvalues of the conserved charges and analyze the spin dynamics. Finally, the paper concludes with a summary of the main results.

\section{ Richardson-Gaudin Models}
Possibly, the simplest non-interacting Hamiltonian one could write for a many-body spin system is given by:
\begin{equation}\label{1}
    H = \sum_{i=1}^{N} H_{i} = \sum_{i=1}^N \epsilon_i S_i^z,
\end{equation}
where $\epsilon_i$ is a free parameter that can be interpreted as a magnetic field in the $z$-direction or energy of the single-particle. The conserved charges associated with this Hamiltonian are simply $Q_{i} = S^{z}_{i}$. This can be easily verified using the condition $[Q_{i}, H] = 0$.

Now, to see how RG models can be used to describe pairing consider the general Hamiltonian for $XXZ$ Richardson-Gaudin models. The Hamiltonian is : 
\begin{equation}\label{pairing}
H = \sum_{i=1}^N \epsilon_i S_i^z + g \sum_{i \neq j}^N \Big( \Gamma_{ij}^x (S_i^+ S_j^- + S_i^- S_j^+) + \Gamma_{ij}^z S_i^z S_j^z \Big),
\end{equation}
where,  $\epsilon_i$ are the single-particle energies (e.g., energy levels of fermions),  $S_i^z, S_i^\pm$ are the spin operators representing particle-hole excitations. $g$ is the coupling strength controlling the interaction between pairs and $\Gamma_{ij}^x, \Gamma_{ij}^z$ are the interaction coefficients that determine the nature of the pairing. For pairing interactions, the term $S_i^+ S_j^-$ describes the annihilation or creation of a pair of particles at sites $i$ and $j$. Physically, this corresponds to two-particles forming a correlated pair. For example, in superconductivity, this describes the formation of Cooper pairs and in nuclear physics, it describes proton-proton or neutron-neutron pairing.  When the coupling strength satisfies $g > 0$, it represents an attractive interaction that favors pairing. Conversely, when $g < 0$, it represents a repulsive interaction that suppresses pairing. Therefore,  the RG models capture the competition between single-particle energies ($\epsilon_i$) and pairing interactions ($g$), providing insight into the emergence of collective behavior. In Fig. 1, we give a pictorial representation of the $XXZ$ RG model.

 \begin{figure}[htbp]
    \centering 
    \begin{tikzpicture}[
        scale=0.9, % Scale for compactness
        spin/.style={draw, circle, minimum size=5mm, fill=blue!20},
        pair/.style={red, thick, <->}, % Use default arrow tips
        coupling/.style={blue, dashed, thick},
        node distance=1.5cm
    ]

        % Draw spins as a linear chain
        \foreach \i/\label in {1/1, 2/2, 3/3, 4/4} {
            \node[spin] (spin\i) at (\i*1.5, 0) {$S_{\label}$};
        }

        % Draw pairing interactions (S_i^+ S_j^- + S_i^- S_j^+)
        \draw[pair] (spin1.north) to[out=60, in=120] node[above, font=\scriptsize] {$S_1^+ S_2^-$} (spin2.north);
        \draw[pair] (spin2.north) to[out=60, in=120] node[above, font=\scriptsize] {$S_2^+ S_3^-$} (spin3.north);
        \draw[pair] (spin3.north) to[out=60, in=120] node[above, font=\scriptsize] {$S_3^+ S_4^-$} (spin4.north);

        % Draw spin-spin interactions (S_i^z S_j^z)
        \draw[coupling] (spin1.south) -- node[below, font=\scriptsize] {$S_1^z S_2^z$} (spin2.south);
        \draw[coupling] (spin2.south) -- node[below, font=\scriptsize] {$S_2^z S_3^z$} (spin3.south);
        \draw[coupling] (spin3.south) -- node[below, font=\scriptsize] {$S_3^z S_4^z$} (spin4.south);

    \end{tikzpicture}
    \caption{Pictorial representation of the $XXZ$ Richardson-Gaudin model. Red arrows indicate pairing interactions ($S_i^+ S_j^-$), while blue dashed lines represent spin-spin interactions ($S_i^z S_j^z$) for the first four spins.}
    \label{xxz}
\end{figure}

A hallmark of Richardson-Gaudin models is their integrability, which arises from the existence of a set of conserved charges $\{Q_i\}$. These conserved charges satisfy the commutation relations:
\begin{equation}
[Q_i, Q_j] = 0, \quad [Q_i, H] = 0.
\end{equation}

In $XXZ$ RG model, the conserved charges take the form:
\begin{equation}
Q_i = S_i^z + g \sum_{j \neq i}^N \Big( \Gamma_{ij}^x (S_i^+ S_j^- + S_i^- S_j^+) + \Gamma_{ij}^z S_i^z S_j^z \Big).
\end{equation}
In the limit $g \to 0$, this reduces to the conserved charges for the non-interacting case. Conversely, when $g \to \infty$, it yields the conserved charges for the $XXZ$ Gaudin magnet \cite{Bethe}. The integrability condition $[Q_{i}, Q_{j}] = 0$ is satisfied if and only if the following conditions hold \cite{Dimo,Faribault}:
\begin{eqnarray}
    \Gamma^{x}_{ij} + \Gamma^{x}_{ji} = 0, \;\;\; \Gamma^{z}_{ij} + \Gamma^{z}_{ji} = 0, \;\; \forall i \neq j, \\
    \Gamma^{x}_{ij}\Gamma^{x}_{jk} - \Gamma^{x}_{ik}(\Gamma^{z}_{ij} + \Gamma^{z}_{jk}) = 0, \;\; \forall i \neq j \neq k.
\end{eqnarray}
The integrability ensures that the eigenstates of the Hamiltonian can be constructed exactly using the Bethe ansatz, which provides a systematic way to solve for the eigenvalues and eigenstates. The rational, trigonometric, and hyperbolic models describe different forms of coupling terms $\Gamma^x_{ij}$ and $\Gamma^z_{ij}$ that arise in integrable systems, such as Richardson-Gaudin models. These models are distinguished by the functional dependence of the coupling terms on the energy differences $\epsilon_i - \epsilon_j$. Notable examples of integrable coupling structures in generalized Gaudin models include the rational, trigonometric, and hyperbolic classes, each distinguished by the functional form of their interaction kernels. In the rational model ($XXX$ model), the coupling terms are inversely proportional to the energy differences between levels:
\begin{equation}
    \Gamma^{x}_{ij} = \frac{1}{\epsilon_i - \epsilon_j}, \quad 
    \Gamma^{z}_{ij} = \frac{1}{\epsilon_i - \epsilon_j}.
\end{equation}
This represents the simplest case, where interactions decay algebraically as the energy separation increases, reflecting a long-range but non-oscillatory character. In contrast, the trigonometric model ($XXZ$ model) introduces periodicity through sine and cotangent functions:
\begin{equation}
    \Gamma^{x}_{ij} = \frac{1}{\sin(\epsilon_i - \epsilon_j)}, \quad 
    \Gamma^{z}_{ij} = \cot(\epsilon_i - \epsilon_j).
\end{equation}
This leads to oscillatory coupling, characteristic of systems with periodic boundary conditions or those defined in terms of angular variables. Finally, the hyperbolic model ($XXZ$ model) features couplings governed by hyperbolic functions:
\begin{equation}
    \Gamma^{x}_{ij} = \frac{1}{\sinh(\epsilon_i - \epsilon_j)}, \quad 
    \Gamma^{z}_{ij} = \coth(\epsilon_i - \epsilon_j).
\end{equation}
These couplings exhibit exponential decay or growth with energy separation, a behavior often associated with systems displaying exponential localization, finite correlation lengths, or effective long-range interactions in non-compact geometries. Together, these models form a fundamental hierarchy of exactly solvable systems, with each class linked to a specific classical $r$-matrix and underlying algebraic structure.

 In trigonometric model,  the coupling terms exhibit periodic behavior due to sine and cotangent functions, with divergences occurring at integer multiples of $\pi$.
In  hyperbolic model, the coupling terms show exponential decay for large $|\epsilon_i - \epsilon_j|$, with $\Gamma^x_{ij}$ decaying rapidly and $\Gamma^z_{ij}$ approaching asymptotic values of $\pm 1$. Singularities are highlighted at $\epsilon_i - \epsilon_j = 0$ for all models.
Each model corresponds to a distinct class of physical systems and reflects different underlying symmetries or interaction mechanisms.\vskip 5mm

The most general member of this integrable hierarchy is the elliptic model (XYZ model), which encompasses the rational, trigonometric, and hyperbolic classes as limiting cases. In this model, the coupling terms are governed by doubly periodic elliptic functions, specifically the Jacobi elliptic functions and their associated Weierstrass or theta-function representations:
\begin{equation}
    \Gamma^{x}_{ij} = \frac{\mathrm{cn}(\epsilon_i - \epsilon_j)}{\mathrm{sn}(\epsilon_i - \epsilon_j)}, \quad
    \Gamma^{y}_{ij} = \frac{\mathrm{dn}(\epsilon_i - \epsilon_j)}{\mathrm{sn}(\epsilon_i - \epsilon_j)}, \quad
    \Gamma^{z}_{ij} = \frac{1}{\mathrm{sn}(\epsilon_i - \epsilon_j)},
\end{equation}
where $\mathrm{sn}$, $\mathrm{cn}$, and $\mathrm{dn}$ are Jacobi elliptic functions with modulus $k \in [0,1]$. This structure introduces two independent periods in the complex plane, reflecting a richer algebraic structures tied to the $r$-matrix of the classical Yang--Baxter equation on a torus. The elliptic model describes systems with modulated long-range interactions and intrinsic anisotropy in all spin components, making it relevant for quantum magnets with competing interactions or systems subject to quasi-periodic potentials. In the limits $k \to 0$ and $k \to 1$, the elliptic couplings reduce to the trigonometric and hyperbolic forms, respectively, thereby unifying all previously discussed classes within a single algebraic framework. Like its descendants, the model exhibits singularities when $\epsilon_i = \epsilon_j$, and its exact solvability arises from a complete set of conserved charges constructed via the algebraic Bethe ansatz. Thus, the elliptic model thus represents the most general exactly solvable central spin or pairing structure in the Richardson-Gaudin family, capturing the full range of integrable coupling behaviors --- from oscillatory and exponential to quasi-periodic --- and forms the basis for understanding a broad class of strongly correlated quantum many-body systems. \vskip 5mm
The most general Bethe root equation is given by:
\begin{equation}
\frac{1}{g} + \sum_{j=1}^N \frac{\Gamma_j}{\epsilon_j - E_\alpha} - \sum_{\beta \neq \alpha}^M \frac{\Lambda_{\alpha\beta}}{E_\beta - E_\alpha} = 0, \quad \alpha = 1, \dots, M.
\end{equation}
This form encompasses a wide variety of integrable models, including Richardson-Gaudin models, Gaudin magnets, and XXZ spin chains. The specific forms of $\Gamma_j$ and $\Lambda_{\alpha\beta}$ depend on the details of the model under consideration. 

The eigenstates of the Richardson-Gaudin Hamiltonian are expressed in terms of Bethe roots $\{E_\alpha\}$ as:
\begin{equation}
|E_1, \dots, E_M\rangle = \prod_{a=1}^M \left( \sum_{i=1}^N \frac{S_i^+}{\epsilon_i - E_a} \right) |\downarrow, \dots, \downarrow\rangle,
\end{equation}
where $S_i^+$ are the spin-raising operators, and the Bethe roots satisfy the coupled algebraic equations \cite{thesis}:
\begin{equation}
\frac{1}{g} + \frac{1}{2}\sum_{j=1}^N \frac{1}{\epsilon_j - E_\alpha} - \sum_{\beta \neq \alpha} \frac{1}{E_\beta - E_\alpha} = 0, \quad \alpha = 1, \dots, M.
\end{equation}
Here, $M$ is the number of pairs, and these equations encode the distribution of pairs across the available single-particle energy levels $\{\epsilon_j\}$. The solutions $\{E_\alpha\}$ correspond to the energies of the paired states. The ground state corresponds to the configuration where all pairs occupy the lowest available energy levels, while excited states arise when pairs are promoted to higher energy levels, creating collective excitations. \vskip 5mm
The algebra underlying the Richardson–Gaudin model constitutes a special case of the more general framework known as the Generalized Gaudin Algebra (GGA), which is defined via operator-valued Lax matrices $L(u)$ that depend on a spectral parameter $u$. These matrices satisfy the following fundamental commutation relation \cite{faddeev,baxter,korepin}:

\begin{equation}
    [L(u), L(v)] = [r(u-v), L(u) \otimes \mathbb{I} + \mathbb{I} \otimes L(v)],
\end{equation}

where,  $u$ and $v$ are spectral parameters,  $r(u-v)$ is the classical $r$-matrix, which encodes the underlying algebraic structure, and $L(u)$ is the Lax matrix, which is  expressed as:
    \begin{equation}
        L(u) = \sum_{i=1}^N \frac{\vec{S}_i \cdot \vec{\sigma}}{u - \epsilon_i},
    \end{equation}
    where,  $\vec{S}_i = (S_i^x, S_i^y, S_i^z)$ are the spin or pseudospin operators at site $i$, $\vec{\sigma} = (\sigma^x, \sigma^y, \sigma^z)$ are the Pauli matrices, and  $\epsilon_i$ are the single-particle energy levels. The transfer matrix, defined as:
\begin{equation}
    T(u) = \text{Tr}(L(u)),
\end{equation}
generates the conserved charges of the system. Specifically, the expansion of $T(u)$ in powers of $u^{-1}$ yields a set of mutually commuting operators:
\begin{equation}
    [T(u), T(v)] = 0, \quad \forall u, v.
\end{equation}
These conserved charges ensure the integrability of the system. The Hamiltonian can then be expressed as a particular linear combination of these charges. The Lax matrix $L(u)$ encodes the dynamics of the system and depends on the spectral parameter $u$. It generalizes the concept of local spin operators to a non-local, operator-valued matrix. The $r$-matrix defines the underlying algebraic structure and ensures the integrability of the system. For the $su(2)$-based Gaudin models, the $r$-matrix is typically the rational solution of the classical Yang-Baxter equation. The trace of the Lax matrix, $T(u)$, generates the conserved charges of the system. The commutativity of $T(u)$ for different values of $u$ ensures the existence of a complete set of commuting operators. In this formalism, the Hamiltonian is derived from the conserved charges generated by the transfer matrix. This construction guarantees that the Hamiltonian is integrable. 
For example, in the RG pairing  Hamiltonian \ref{pairing},  the interaction coefficients $\Gamma_{ij}^x$ and $\Gamma_{ij}^z$ are determined by the structure of the $r$-matrix. In this case, the Lax operator at site $i$ is given by:
\begin{equation}
L_i(u) = \frac{1}{u - \epsilon_i}
\begin{pmatrix}
S_i^z & S_i^- \\
S_i^+ & -S_i^z
\end{pmatrix}.
\end{equation}
and the total Lax operator for the system is:
\begin{equation}
L(u) = \sum_{i=1}^N L_i(u).
\end{equation}
The classical $r$-matrix is defined as:
\begin{equation}
r(u-v) = \frac{\vec{\sigma} \otimes \vec{\sigma}}{u - v},
\end{equation}
where $\vec{\sigma} \otimes \vec{\sigma}$ represents the tensor product of Pauli matrices:
\begin{equation}
\vec{\sigma} \otimes \vec{\sigma} = \sigma^x \otimes \sigma^x + \sigma^y \otimes \sigma^y + \sigma^z \otimes \sigma^z.
\end{equation}
The transfer matrix is obtained by taking the trace of the Lax operator:
\begin{equation}
T(u) = \text{Tr}(L(u)) = \sum_{i=1}^N \frac{S_i^z}{u - \epsilon_i}.
\end{equation}
This generates the conserved charges of the system.
The generalized Gaudin algebra (GGA) is defined by the commutation relations \cite{GGA}:

\begin{align}\label{algebra1}
    \nonumber [S_x(u), S_y(v)] &= i \big(Y(u, v)S_z(u) - X(u, v)S_z(v)\big), \\ \nonumber 
    [S_y(u), S_z(v)] &= i \big(Z(u, v)S_x(u) - Y(u, v)S_x(v)\big), \nonumber \\
    [S_z(u), S_x(v)] &= i \big(X(u, v)S_y(u) - Z(u, v)S_y(v)\big), \\  \nonumber
    [S_\alpha(u), S_\alpha(v)] &= 0, \quad \alpha = x, y, z.
\end{align}
where $ u, v \in \mathbb{C} $. This represents an infinite-dimensional Lie algebra that strongly resembles the $ su(2) $ algebra. It is defined by three functions: $ X(u, v) $, $ Y(u, v) $, and $ Z(u, v) $. 
For the  RG pairing Hamiltonian \ref{pairing},  GGA is defined by a set of operators $\{S_i^z, S_i^+, S_i^-\}$ at each site $i = 1, \dots, N$, satisfying the following  commutation relations:
\begin{align}\label{algebra2}
    [S_i^z, S_j^\pm] &= \pm \delta_{ij} S_i^\pm, \nonumber \\
    [S_i^+, S_j^-] &= 2 \delta_{ij} S_i^z, \nonumber  \\
    [S_i^z, S_j^z] &= 0.
\end{align}

\section{Results}
In this section, we present our main results on $\mathcal{PT}$-symmetric Richardson--Gaudin models. We begin by constructing a consistent $\mathcal{PT}$-symmetric framework for spin-$\tfrac{1}{2}$ systems, where parity is defined as $\mathcal{P} = \prod_i \sigma_i^z$ and time-reversal flips only the $y$-component of spin. We show that the resulting non-Hermitian Hamiltonian remains integrable and derive its Hermitian counterpart via an explicit similarity transformation, identifying the metric operator that defines the physical inner product. Next, we analyze the spectral properties of the conserved charges, demonstrating that eigenvalues are either real or form complex conjugate pairs—characteristic of $\mathcal{PT}$ symmetry—and reveal partial symmetry breaking wherein low-energy states remain in the unbroken phase while higher-energy states transition to the broken phase. Finally, we provide exact analytical expressions for spin dynamics in both phases, showing coherent oscillations in the unbroken regime and exponentially modulated behavior in the broken regime. 

\subsection{$\mathcal{PT}$-symmetric Richardson-Gaudin Hamiltonians}
To construct $\mathcal{PT}$-symmetric Richardson--Gaudin models, one must first define appropriate parity ($\mathcal{P}$) and time-reversal ($\mathcal{T}$) transformations. The parity operator acts as a reflection in the $xy$-plane, implementing the transformation
\begin{equation}
(S^{x}_{i}, S^{y}_{i}, S^{z}_{i}) \mapsto (-S^{x}_{i}, -S^{y}_{i}, S^{z}_{i}).
\end{equation}
For spin-$\tfrac{1}{2}$ systems, this is realized by the unitary operator $\mathcal{P}_i = \sigma_i^z$, so that the total parity operator is
\begin{equation}
\mathcal{P} = \prod_{i=1}^{N} \sigma_i^z.
\end{equation}

The time-reversal operator $\mathcal{T}$ is antiunitary. In the context of $\mathcal{PT}$-symmetric spin models, we adopt the convention that $\mathcal{T}$ acts as complex conjugation $\mathcal{K}$ combined with a spin rotation such that only the $y$-component of spin changes sign \cite{Fring}. Specifically,
\begin{equation}
\mathcal{T} S^{x}_i \mathcal{T}^{-1} = S^{x}_i, \quad
\mathcal{T} S^{y}_i \mathcal{T}^{-1} = -S^{y}_i, \quad
\mathcal{T} S^{z}_i \mathcal{T}^{-1} = S^{z}_i.
\end{equation}

A Hamiltonian $H$ is said to be $\mathcal{PT}$-symmetric if it commutes with the combined antiunitary operator $\mathcal{PT}$, i.e.,
\begin{equation}
[\mathcal{PT}, H] = 0.
\end{equation}
When this symmetry is unbroken, the eigenvalues of $H$ remain real despite $H$ being non-Hermitian. In this regime, the right and left eigenvectors satisfy the biorthogonal eigenvalue equations
\begin{align}
    H |\phi_n\rangle &= E_n |\phi_n\rangle, \\
    H^\dagger |\psi_n\rangle &= E_n^* |\psi_n\rangle,
\end{align}
and form a complete biorthonormal basis:
\begin{align}
    \langle \psi_n | \phi_m \rangle &= \delta_{nm}, \\
    \sum_n |\phi_n\rangle \langle \psi_n| &= \mathbb{I}.
\end{align}
In the unbroken $\mathcal{PT}$-symmetric phase, $E_n \in \mathbb{R}$, so $E_n^* = E_n$.

The parity operator $\mathcal{P}$ is Hermitian and unitary, satisfying $\mathcal{P}^\dagger = \mathcal{P}$ and $\mathcal{P}^2 = \mathbb{I}$. For a $\mathcal{PT}$-symmetric Hamiltonian, one has the relation
\begin{equation}
H^\dagger = \mathcal{P} H \mathcal{P},
\end{equation}
which follows from $[\mathcal{PT}, H] = 0$ and the antiunitarity of $\mathcal{T}$.

In the unbroken phase, the $\mathcal{PT}$ operator maps right eigenvectors to themselves up to a phase:
\begin{equation}
\mathcal{PT} |\phi_n\rangle = \lambda_n |\phi_n\rangle, \quad |\lambda_n| = 1.
\end{equation}
One can then define a positive-definite metric operator $\rho = \sum_n |\phi_n\rangle \langle \phi_n|$ (or equivalently $\rho = \mathcal{PC}$ in the standard $\mathcal{CPT}$ framework), which satisfies
\begin{equation}
H^\dagger = \rho H \rho^{-1}.
\end{equation}
This metric operator defines a new inner product
\begin{equation}
\langle \phi | \psi \rangle_{\rho} = \langle \phi | \rho \psi \rangle,
\end{equation}
with respect to which $H$ is self-adjoint. Consequently, there exists a similarity transformation $\eta = \sqrt{\rho}$ such that the Hermitian counterpart of $H$ is given by
\begin{equation}\label{hermitian}
    h = \eta H \eta^{-1} = h^\dagger.
\end{equation}
This construction applies to any observable $O$ in the theory: its Hermitian counterpart is $o = \eta O \eta^{-1}$.

\vskip 5mm
Consider the Hamiltonian \ref{pairing} with  deformation $\Gamma^{x}_{ij}\rightarrow \hat{\Gamma}^{x}_{ij}= i\gamma_{ij}$ while $ \hat{\Gamma}^{z}_{ij}\in \mathbb{R}$. The resulting Hamiltonian is 

\begin{equation}\label{nhpairing}
H_{\mathrm{RG}} = \sum_{i=1}^N \epsilon_i S_i^z + g \sum_{i \neq j}^N \Big( \hat{\Gamma}_{ij}^x (S_i^+ S_j^- + S_i^- S_j^+) +  \hat{\Gamma}_{ij}^z S_i^z S_j^z \Big),
\end{equation}

Although the interacting term is non-Hermitian, the Hamiltonian $H_{\text{RG}}$ can still be $\mathcal{PT}$-symmetric since  it satisfies the condition:
\begin{equation}
H_{\text{RG}}^\dagger = \mathcal{PT}H_{\text{RG}} (\mathcal{PT})^{-1}.
\end{equation}

This can be proven by applying the combined $\mathcal{PT}$ transformation, which constitutes a generalized antiunitary symmetry and induces the following action:
\begin{align}
\mathcal{PT} S_i^z (\mathcal{PT})^{-1} &= S_i^z, \\
\mathcal{PT} S_i^+ (\mathcal{PT})^{-1} &= -S_i^-, \\
\mathcal{PT} S_i^- (\mathcal{PT})^{-1} &= -S_i^+.
\end{align}

Next, let us find the hermitian counterpart of the RG pairing Hamiltonian \ref{nhpairing} using \ref{hermitian}.  The metric operator $\rho$ is constructed to ensure that the transformed Hamiltonian becomes Hermitian. For $\mathcal{PT}$-symmetric systems, $\eta$ can often be chosen as a diagonal operator in the spin basis. A common choice for $\eta$ is:
\begin{equation}
\eta = e^{-Q/2},
\end{equation}
where $Q$ is an anti-Hermitian operator that encodes the deviation from Hermiticity. In this case, $Q$ is related to the imaginary part of $\Gamma_{ij}^x$.

For simplicity, assume $Q$ acts locally on each spin site as:
\begin{equation}
Q = \sum_{i=1}^N q_i S_i^z,
\end{equation}
where $q_i$ are real parameters determined by $\hat{\Gamma}_{ij}^x$.

Thus:
\begin{equation}
\eta = e^{-\frac{1}{2} \sum_{i=1}^N q_i S_i^z}.
\end{equation}
The metric operator is then:
\begin{equation}
\rho = \eta^\dagger \eta = e^{-\sum_{i=1}^N q_i S_i^z}.
\end{equation}

The Hermitian counterpart $h_{\text{RG}}$ is obtained via the similarity transformation:
\begin{equation}
h_{\text{RG}} = \eta\;  H_{\text{RG}}\;  \eta^{-1}.
\end{equation}
The terms $\epsilon_i S_i^z$ are unaffected by $\eta$, because $\eta$ commutes with $S_i^z$. The raising and lowering operators $S_i^+$ and $S_i^-$ transform under $\eta$ as: $
\eta S_i^+ \eta^{-1} = e^{-q_i / 2} S_i^+, \quad \eta S_i^- \eta^{-1} = e^{q_i / 2} S_i^-.
$
The commutator of $ Q $ with $ S_i^+ $ is:
\begin{equation}
[Q, S_i^+] = \left[\sum_{k=1}^N q_k S_k^z, S_i^+\right] = q_i [S_i^z, S_i^+] = q_i S_i^+,
\end{equation}
since $[S_i^z, S_i^+] = S_i^+$. 
Using the Baker-Campbell-Hausdorff (BCH) identity:
\begin{equation}
e^{-Q/2} S_i^+ e^{Q/2} = S_i^+ + \frac{1}{1!}(-Q/2)S_i^+ + \frac{1}{2!}(-Q/2)^2 S_i^+ + \dots
\end{equation}

Substituting $[Q, S_i^+] = q_i S_i^+$, this simplifies to $ e^{-Q/2} S_i^+ e^{Q/2} = e^{-q_i / 2} S_i^+.$ Similarly,$
e^{-Q/2} S_i^- e^{Q/2} = e^{q_i / 2} S_i^-.$
Thus, the term $S_i^+ S_j^-$ transforms as: $\eta (S_i^+ S_j^-) \eta^{-1} = e^{-(q_i - q_j)/2} S_i^+ S_j^-.
$ Similarly, the term $S_i^- S_j^+$ transforms as:
$
\eta (S_i^- S_j^+) \eta^{-1} = e^{(q_i - q_j)/2} S_i^- S_j^+.
$
and 
$
\eta (S_i^- S_j^+) \eta^{-1} = e^{(q_i - q_j)/2} S_i^- S_j^+.
$
Therefore, the transformed Hamiltonian $h$ becomes:
\begin{equation}
h_{\text{RG}} = \sum_{i=1}^N \epsilon_i S_i^z + g \sum_{i \neq j}^N \Big( \tilde{\Gamma}_{ij}^x (S_i^+ S_j^- + S_i^- S_j^+) + \hat{\Gamma}_{ij}^z S_i^z S_j^z \Big),
\end{equation}
where:
\begin{equation}
\tilde{\Gamma}_{ij}^x = \hat{\Gamma}_{ij}^x e^{q_i - q_j}.
\end{equation}
Since $\hat{\Gamma}_{ij}^x = i\gamma_{ij}$, the new coupling constants $\tilde{\Gamma}_{ij}^x$ must be chosen such that $\tilde{\Gamma}_{ij}^x$ is real. This ensures that $h$ is Hermitian.
The Hermitian counterpart $h_{\text{RG}}$ is:
\begin{equation}\label{trans}
h_{\text{RG}} = \sum_{i=1}^N \epsilon_i S_i^z + g \sum_{i \neq j}^N \Big( \text{Re}(\tilde{\Gamma}_{ij}^x) (S_i^+ S_j^- + S_i^- S_j^+) + \hat{\Gamma}_{ij}^z S_i^z S_j^z \Big),
\end{equation}
where $
\tilde{\Gamma}_{ij}^x$ does not include a division by 2 because the full exponential factor $e^{q_i - q_j}$ is already incorporated into the definition of $\tilde{\Gamma}_{ij}^x$. The factor of $1/2$ appears in the intermediate steps when transforming individual operators ($S_i^+$ and $S_i^-$). The reason why the transformed coupling constant $\tilde{\Gamma}_{ij}^x$ is written as the same in \ref{trans} for both pairing terms $S_i^+ S_j^-$ and $S_i^- S_j^+$ lies in the symmetry of the Hamiltonian under the transformation induced by $\eta$. The parameters $q_i$ are determined by requiring $\text{Im}(\tilde{\Gamma}_{ij}^x) = 0$, ensuring that $h$ is Hermitian. Finally, it is worth noting that the algebra of the $\mathcal{PT}$-symmetric pairing  Richardson--Gaudin model satisfies~ \ref{algebra2}.  Consequently, the algebraic structure of the commutators naturally extends to the $\mathcal{PT}$-symmetric regime.
\subsection{Hamiltonian Spectrum}
For the sake of universality, we consider the case of non-Hermitian $XYZ$ Richardson-Gaudin models for spin-$1/2$ particles subjected to an arbitrary magnetic field. The magnetic field components $B_x$ and $B_y$, as well as the coupling constants $\hat{\Gamma}^x_{ik}$ and $\hat{\Gamma}^y_{ik}$, are taken to be complex. The $\mathcal{PT}$-symmetric  Richardson-Gaudin pairing Hamiltonian represents a special case within this more general framework.
The general form for the conserved charges that are quadratic in Pauli matrices is \cite{Dimo}
\begin{widetext}
\begin{eqnarray}
    Q_{i} &=&  B^{x}_{i} S^{x}_{i} + B^{y}_{i} S^{y}_{i} + B^{z}_{i} S^{z}_{i} + \sum_{k \neq i}^{N} \left( \hat{\Gamma}^{x}_{ik} S^{x}_{i} S^{x}_{k} + \hat{\Gamma}^{y}_{ik} S^{y}_{i} S^{y}_{k} + \hat{\Gamma}^{z}_{ik} S^{z}_{i} S^{z}_{k} \right),
\end{eqnarray}
\end{widetext}
where $B_{i}^{x},B_{i}^{y}, \hat{\Gamma}_{ik}^{x},$ and $\hat{\Gamma}_{ik}^{y} \in \mathbb{C}$ are complex-valued quantities. 
The metric operators $\eta = e^{-\frac{1}{2} \sum_{i=1}^N q_i S_i^z}$, $\rho = e^{-\sum_{i=1}^N q_i S_i^z}$, where $q_i$ are real parameters determined by the system's properties, are defined as in the previous section.

The Hermitian counterpart $\tilde{Q}_i$ of $Q_i$ is obtained through the similarity transformation:
\begin{equation}
    \tilde{Q}_i = \eta^{-1} Q_i \eta.
\end{equation}
Under this transformation, the spin operators $S_i^\alpha$, with $\alpha \in \{x, y, z\}$, transform as:
\begin{align}
    \eta^{-1} S_i^\alpha \eta &= e^{q_i} S_i^\alpha, \quad \alpha = x, y, \\
    \eta^{-1} S_i^z \eta &= S_i^z.
\end{align}
    
Thus, $\tilde{Q}_i$ is:
\begin{widetext}
\begin{equation}
\tilde{Q}_i = e^{q_i} B_i^x S_i^x + e^{q_i} B_i^y S_i^y + B_i^z S_i^z + \sum_{k \neq i}^N \left( e^{q_i + q_k} \hat{\Gamma}_{ik}^x S_i^x S_k^x + e^{q_i + q_k} \hat{\Gamma}_{ik}^y S_i^y S_k^y + \hat{\Gamma}_{ik}^z S_i^z S_k^z \right).
\end{equation}
\end{widetext}
The parameters $q_i$ are determined by requiring that the transformed operator $\tilde{Q}_i$ is Hermitian. Since $B_i^x$, $B_i^y$, $\hat{\Gamma}_{ik}^x$, and $\hat{\Gamma}_{ik}^y$ are purely imaginary, the exponential factors $e^{q_i}$ and $e^{q_i + q_k}$ must cancel the imaginary parts. This leads to the condition:
\begin{equation}
e^{q_i} B_i^x, \, e^{q_i} B_i^y, \, e^{q_i + q_k} \hat{\Gamma}_{ik}^x, \, e^{q_i + q_k} \hat{\Gamma}_{ik}^y \in \mathbb{R}.
\end{equation}

For example, if $B_i^x = i b_i^x$ (where $b_i^x \in \mathbb{R}$), then $e^{q_i} B_i^x = e^{q_i} i b_i^x \in \mathbb{R}$ implies $ e^{q_i} = |b_i^x|^{-1}.$
 Similarly, $q_i + q_k$ is determined by $\hat{\Gamma}_{ik}^x$ and $\hat{\Gamma}_{ik}^y$. 
  \vskip 5mm
For any permutation of $(\alpha,\beta,\gamma)$ from the set $\{x,y,z\}$, the integrability conditions are :
\begin{eqnarray}\label{integra1}
    \hat{\Gamma}^{\beta}_{ij}B^{\alpha}_{j}+ \hat{\Gamma}^{\alpha}_{ji}B^{\alpha}_{i}=0, \; \forall i\neq j, \\ \label{integra2}
    \hat{\Gamma}^{\alpha}_{ik}\hat{\Gamma}^{\beta}_{jk}-\hat{\Gamma}^{\alpha}_{ij}\hat{\Gamma}^{\gamma}_{jk}- \hat{\Gamma}^{\beta}_{ji}\hat{\Gamma}^{\gamma}_{ik}=0 , \;\; \forall i\neq j\neq k, 
\end{eqnarray}
where $\hat{\Gamma}^{\alpha}_{ij}=(\hat{\Gamma}^{x}_{ij},\hat{\Gamma}_{ij}^{y}, \hat{\Gamma}_{ij}^{z})=(i\Gamma^{x}_{ij},i\Gamma^{y}_{ij},\Gamma^{z}_{ij})$ and $B^{\alpha}_{i}=(iB^{x}_{i},iB^{y}_{i},B^{z}_{i})$. 
In analogy with \cite{Dimo}, one can derive quadratic operator identities for spin-$1/2$ Richardson--Gaudin models, which relate $Q_i^2$ to linear combinations of the conserved charges $Q_j$. 
\begin{equation}
Q_i^2 = \sum_{j\ne i} C_{ij} Q_j + K_i, 
\end{equation}
with 
\begin{eqnarray}
    K_i =  \left(\sum_{\alpha} (B^\alpha_i)^2 + \sum_{\alpha} \sum_{k\ne i }^N \left( \hat{\Gamma}^{\alpha}_{i k} \right)^2\right), \\
    C_{ik} = 
\begin{cases} 
2 \frac{B_i^\alpha \hat{\Gamma}_{ik}^\alpha}{B_k^\alpha}, & \text{(from linear terms)}, \\
-2 \frac{ \hat{\Gamma}{ik}^\beta \hat{\Gamma}_{ik}^\gamma}{\hat{\Gamma}_{ki}^\alpha}, & \text{(from quadratic terms)}.
\end{cases}
\end{eqnarray}
These relations hold even in the presence of complex magnetic field vectors and coupling constants, thereby extending the integrable structure to non-Hermitian settings.
The linear terms (linear in Pauli operators) are the  self-interaction of spins at site $i$, represented by the term $\vec{B}_i \cdot \vec{S}_i$. 
This linear term arises from the coupling of the magnetic field $\vec{B}_i$ with the spin operator $\vec{S}_i$, and it primarily contributes 
to the constant term $K_i$. Therefore, the linear nature of $\vec{B}_i \cdot \vec{S}_i$ ensures that it does not directly influence the coefficients $C_{ij}$, but rather 
contributes solely to $K_i$.
The term $\hat{\Gamma}_{ik}^\alpha S_i^\alpha S_k^\alpha$ describes the interaction between spins at site $i$ and spins at other sites $k$. 
This interaction is mediated by the coupling strength $\hat{\Gamma}_{ik}^\alpha$, which couples the $\alpha$-component of the spin operator $S_i^\alpha$ 
at site $i$ with the corresponding $\alpha$-component of the spin operator $S_k^\alpha$ at site $k$.
 
These expressions ensure that $C_{ik}$ satisfies all integrability conditions \ref{integra1} and \ref{integra2}. In \cite{Nadai}, the authors  chose to assign $ Q_0$ ($Q_{1}$ in our notation)  as the Hamiltonian due to its physical significance. However, the same properties would hold true for any other Hamiltonian formed from the conserved charges, as they would all exhibit the same eigenstates. In the context of this work, we do not specify a precise Hamiltonian because it is not necessary. The operators $ Q_1 $ and $ Q_i $ all commute with one another, meaning they share the same set of eigenvectors. As a result, whether we  choose $ H = Q_1 $ or any linear (or even non-linear) combination:
\begin{equation}
    H = \sum_{i}^N \alpha_i Q_i
\end{equation}
of the conserved charges, the resulting Hamiltonians will always share the same eigenvectors. Of course, the eigenvalues of $ H $ will differ depending on the specific choice of $ H $. 
The integrability conditions for the conserved charges $Q_i$ are derived by demanding mutual commutativity:

\begin{equation}
    [Q_i, Q_j] = 0 \quad \forall\, i \neq j.
\end{equation}

This condition imposes stringent algebraic constraints on the magnetic field components $B_i^\alpha$ and the interaction couplings $\hat{\Gamma}_{ij}^\alpha$. Following the approach of \cite{Faribault, Dimo}, we assume a rational dependence of the couplings on a set of inhomogeneity parameters $\epsilon_i$, inspired by the Gaudin model structure.

The computation of  commutator $[Q_i, Q_j]$ generates terms involving products of spin operators at sites $i$, $j$, and any third site $k$. For these to cancel identically for all spin configurations, the coefficients must satisfy functional equations.  After algebraic calculations,  we write the integrability constraints for an arbitrary magnetic field as: 
\begin{eqnarray}
    B^{x}_{i}= \frac{i\delta}{\sqrt{\alpha_{x}\epsilon_{i}+\beta_{x}}}, \\
    B^{y}_{i}= \frac{i\lambda}{\sqrt{\alpha_{y}\epsilon_{i}+\beta_{y}}}, \\ B^{z}_{i}=1
\end{eqnarray}

\begin{eqnarray}
  \hat{ \Gamma}^{x}_{ij}= ig \frac{\sqrt{(\alpha_{x}\epsilon_{i}+\beta_{x})(\alpha_{y}\epsilon_{j}+\beta_{y})}}{\epsilon_{i}-\epsilon_{j}}, \\
    \hat{\Gamma}^{y}_{ij}= ig \frac{\sqrt{(\alpha_{y}\epsilon_{i}+\beta_{y})(\alpha_{x}\epsilon_{j}+\beta_{x})}}{\epsilon_{i}-\epsilon_{j}},\\
    \hat{\Gamma}^{z}_{ij}= g \frac{\sqrt{(\alpha_{x}\epsilon_{j}+\beta_{x})(\alpha_{y}\epsilon_{j}+\beta_{y})}}{\epsilon_{i}-\epsilon_{j}}
\end{eqnarray}

\begin{figure}
\centering 
\includegraphics[scale=0.45]{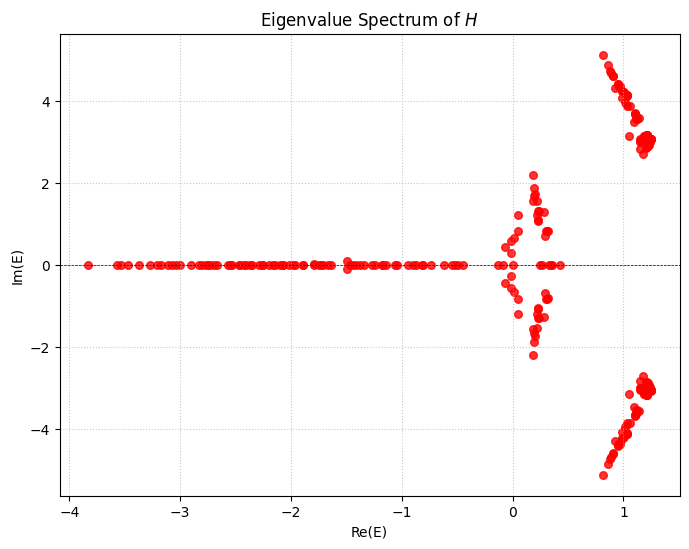}
\caption{
Eigenvalue spectrum of the $\mathcal{PT}$-symmetric Hamiltonian $H = Q_1$ for $N=8$ spin-$\tfrac{1}{2}$ particles. The transverse magnetic fields and couplings are purely imaginary, while longitudinal terms are real, ensuring $\mathcal{PT}$ symmetry. Eigenvalues are either strictly real (on the horizontal axis) or occur in complex conjugate pairs (symmetric about the real axis), as required by $\mathcal{PT}$ symmetry. Notably, all negative-energy states remain real, indicating that the low-energy sector resides in the unbroken $\mathcal{PT}$-symmetric phase.
}
\label{fig:spectrum}
\end{figure}

The explicit forms of the magnetic field components \(B_i^\alpha\) and coupling constants \(\hat{\Gamma}_{ij}^\alpha\) are fully determined by the integrability conditions \([Q_i, Q_j] = 0\). A detailed derivation of these expressions is provided in Appendix~\ref{app:derivation}. \vskip 5mm
To verify the spectral properties of our $\mathcal{PT}$-symmetric Richardson--Gaudin model, we numerically diagonalize the first conserved charge $Q_1$, which we identify as the Hamiltonian $H = Q_1$. We construct $H$ for $N=8$ spin-$\tfrac{1}{2}$ particles with randomly sampled inhomogeneity parameters $\{\epsilon_i\}$. The transverse magnetic field components and coupling constants are chosen to be purely imaginary ($B_i^{x,y}, \hat{\Gamma}_{ij}^{x,y} \in i\mathbb{R}$), while longitudinal terms remain real, ensuring $\mathcal{PT}$ symmetry. The resulting $256 \times 256$ Hamiltonian matrix is diagonalized exactly, and its eigenvalue spectrum is shown in Fig.~\ref{fig:spectrum}. As expected for a $\mathcal{PT}$-symmetric system, all eigenvalues are either strictly real or form complex conjugate pairs. Notably, the negative-energy sector remains entirely real, indicating that low-lying states reside in the unbroken $\mathcal{PT}$-symmetric phase, while higher-energy states exhibit spontaneous symmetry breaking through complex conjugate pairs. This asymmetric spectral structure reflects the dominance of Hermitian longitudinal interactions in the ground-state manifold, consistent with the physical interpretation of partial $\mathcal{PT}$ symmetry breaking in many-body systems.

The Hamiltonian \( H = Q_1 \) possesses a rich spectral structure governed by \(\mathcal{PT}\) symmetry. Since \( H \) commutes with the antiunitary operator \(\mathcal{PT}\), i.e., \([\mathcal{PT}, H] = 0\), its eigenvalues are either entirely real or occur in complex conjugate pairs. The transition between these two regimes occurs at exceptional points (EPs)—branch-point singularities in the complex energy plane where both eigenvalues and their corresponding eigenvectors coalesce.

To analyze this transition, we introduce a dimensionless non-Hermiticity parameter $\gamma \in \mathbb{R}_{\geq 0}$ that scales the anti-Hermitian part of $H$:
\begin{equation}
    H(\gamma) = H_0 + i\gamma V,
\end{equation}
where $H_0$ is Hermitian (the longitudinal Ising-like part) and $V$ is Hermitian (the transverse coupling strength). Explicitly, in our model:
\begin{align}
    H_0 &= B_z S_1^z + \sum_{k \neq 1} \hat{\Gamma}_{1k}^z S_1^z S_k^z, \\
    V &= \delta \frac{S_1^x}{\sqrt{\alpha_x \epsilon_1 + \beta_x}} + \lambda \frac{S_1^y}{\sqrt{\alpha_y \epsilon_1 + \beta_y}} \nonumber \\
    &\quad + g \sum_{k \neq 1} \Bigg[ 
        \frac{\sqrt{(\alpha_x \epsilon_1 + \beta_x)(\alpha_y \epsilon_k + \beta_y)}}{\epsilon_1 - \epsilon_k} S_1^x S_k^x \nonumber \\
    &\qquad\qquad\quad + \frac{\sqrt{(\alpha_y \epsilon_1 + \beta_y)(\alpha_x \epsilon_k + \beta_x)}}{\epsilon_1 - \epsilon_k} S_1^y S_k^y
    \Bigg].
\end{align}
When $\gamma = 0$, $H$ is Hermitian and all eigenvalues are real (the $\mathcal{PT}$-exact phase). As $\gamma$ increases, level repulsion in the real spectrum weakens until, at critical values $\gamma = \gamma_c^{(n)}$, pairs of eigenvalues collide and bifurcate into the complex plane—marking the onset of the $\mathcal{PT}$-broken phase.

At an exceptional point $\gamma = \gamma_c$, the Hamiltonian becomes defective: for some eigenvalue $E_c$, the algebraic multiplicity exceeds the geometric multiplicity, i.e.,
\begin{equation}
    \dim \ker(H - E_c \mathbb{I}) < \text{mult}(E_c).
\end{equation}
This manifests as the coalescence of both eigenvalues and eigenvectors:
\begin{equation}
    \lim_{\gamma \to \gamma_c} E_i(\gamma) = \lim_{\gamma \ \to \gamma_c} E_j(\gamma) = E_c, \quad
    \lim_{\gamma \to \gamma_c} |\psi_i(\gamma)\rangle = \lim_{\gamma \to \gamma_c} |\psi_j(\gamma)\rangle.
\end{equation}
The EPs are square-root branch points: near $\gamma_c$, the eigenvalues behave as
\begin{equation}
    E_{\pm}(\gamma) \approx E_c \pm c \sqrt{\gamma - \gamma_c},
\end{equation}
for some constant $c \in \mathbb{C}$, confirming their non-analytic nature.

In the many-body context, the $\mathcal{PT}$-symmetry breaking is typically partial: low-energy states (dominated by $H_0$) remain in the exact phase ($E \in \mathbb{R}$) even when higher-energy states enter the broken phase ($E \in \mathbb{C} \setminus \mathbb{R}$). This hierarchy arises because the matrix elements of $V$ between low-lying states are suppressed, making them less susceptible to non-Hermitian perturbations. Consequently, the ground state often remains real up to a critical $\gamma_g > \gamma_c^{(n)}$ for excited states.

\subsection{Spin Dynamics}
The time evolution of local spin observables in the $\mathcal{PT}$-symmetric Richardson--Gaudin model can be solved exactly due to the integrability of the conserved charges $Q_i$. For the Hamiltonian $H = Q_1$, the Heisenberg equation of motion for the spin operator $S_j^z(t)$ reads:
\begin{equation}
    \frac{d}{dt} S_j^z(t) = i [H, S_j^z(t)].
\end{equation}
However, a more powerful approach exploits the fact that $H$ shares eigenstates with all conserved charges $Q_i$. Let $\{ |\phi_n\rangle \}_{n=1}^{2^N}$ denote the right eigenvectors of $H$ with eigenvalues $E_n$, and $\{ |\psi_n\rangle \}$ the corresponding left eigenvectors, satisfying the biorthonormality condition $\langle \psi_m | \phi_n \rangle = \delta_{mn}$.

For an initial state $|\psi(0)\rangle = |\downarrow\downarrow\cdots\downarrow\rangle = \bigotimes_{i=1}^N |1\rangle_i$, the time-evolved state is:
\begin{equation}
    |\psi(t)\rangle = e^{-iHt} |\psi(0)\rangle = \sum_{n=1}^{2^N} c_n e^{-iE_n t} |\phi_n\rangle,
\end{equation}
where $c_n = \langle \psi_n | \psi(0) \rangle$. The physical expectation value of $S_j^z(t)$, computed in the $\mathcal{PT}$-symmetric inner product, is:
\begin{equation}
    \langle S_j^z(t) \rangle_\rho = \frac{\langle \psi(t) | \rho S_j^z | \psi(t) \rangle}{\langle \psi(t) | \rho | \psi(t) \rangle},
\end{equation}
with $\rho = \mathcal{PC}$ the positive-definite metric operator.

Using the resolution of identity $\sum_n |\phi_n\rangle \langle \psi_n| = \mathbb{I}$ and the relation $\rho |\phi_n\rangle = |\psi_n\rangle$ (which holds in the unbroken $\mathcal{PT}$ phase), the numerator becomes:
\begin{align}
    \langle \psi(t) | \rho S_j^z | \psi(t) \rangle 
    &= \sum_{m,n} c_m^* c_n e^{i(E_m - E_n)t} \langle \psi_m | S_j^z | \phi_n \rangle.
\end{align}
Similarly, the denominator is:
\begin{equation}
    \langle \psi(t) | \rho | \psi(t) \rangle = \sum_{n} |c_n|^2.
\end{equation}
In the unbroken $\mathcal{PT}$ phase, where all $E_n \in \mathbb{R}$, this yields a closed-form expression:
\begin{equation}
    \langle S_j^z(t) \rangle_\rho = \frac{
        \sum_{m,n} c_m^* c_n \langle \psi_m | S_j^z | \phi_n \rangle e^{i(E_m - E_n)t}
    }{
        \sum_{n} |c_n|^2
    }
   \end{equation}

For the specific initial state $|\psi(0)\rangle = |\downarrow\downarrow\cdots\downarrow\rangle$, the coefficients simplify to $c_n = \langle \psi_n | \downarrow\cdots\downarrow \rangle$. In the weak-coupling limit ($g \to 0$), the Hamiltonian reduces to $H \approx B_z S_1^z + \sum_{k \neq 1} \hat{\Gamma}_{1k}^z S_1^z S_k^z$, and the dynamics become purely oscillatory:
\begin{equation}
    \langle S_j^z(t) \rangle_\rho \approx 
    \begin{cases}
        -\frac{1}{2} \cos(\Omega t), & j = 1, \\
        -\frac{1}{2}, & j \neq 1,
    \end{cases}
\end{equation}
where $\Omega = 2|B_z + \sum_{k \neq 1} \hat{\Gamma}_{1k}^z|$ is the Rabi frequency.

In the strong-coupling regime, when $\mathcal{PT}$ symmetry is broken and eigenvalues form complex conjugate pairs $E_n = E_n^r \pm i E_n^i$, the dynamics acquire exponential envelopes:
\begin{equation}
    \langle S_j^z(t) \rangle_\rho \sim \sum_{n} A_n e^{-2 E_n^i t} \cos(2 E_n^r t + \phi_n),
\end{equation}
reflecting the effective gain/loss in the system. This analytical structure—coherent oscillations in the unbroken phase and damped modes in the broken phase—is a universal signature of $\mathcal{PT}$-symmetric quantum dynamics.
We emphasize that all spin expectation values are computed using the physical $\mathcal{CPT}$-inner product, i.e.,  
\[
\langle S_j^z(t) \rangle_{\mathcal{CPT}} = \frac{\langle \psi(t) | \mathcal{CPT} \, S_j^z \, | \psi(t) \rangle}{\langle \psi(t) | \mathcal{CPT} | \psi(t) \rangle},
\]  
which ensures real-valued, norm-conserving dynamics consistent with the probabilistic interpretation of $\mathcal{PT}$-symmetric quantum mechanics. In our model, the metric operator $\rho = \mathcal{PC}$ implements this inner product, so the denominator remains constant in time (equal to 1 for normalized initial states), and the numerator yields physically meaningful observables.
\section{Conclusion}
In this work, we have constructed a family of $\mathcal{PT}$-symmetric Richardson--Gaudin models by deforming the closed integrable system through complex-valued magnetic fields and coupling constants. By defining parity as $\mathcal{P} = \prod_i \sigma_i^z$ and adopting a time-reversal operator that flips only the $y$-component of spin, we established a consistent $\mathcal{PT}$-symmetric framework for spin-$\tfrac{1}{2}$ systems. Crucially, our model arises from a direct deformation of the Hermitian Richardson--Gaudin Hamiltonian, in contrast to open-system treatments that introduce dissipation via Lindblad operators.
We demonstrated that the resulting non-Hermitian Hamiltonian remains integrable, with conserved charges that satisfy mutual commutativity under generalized integrability conditions. Through an explicit similarity transformation, we derived the Hermitian counterpart of the system, confirming its pseudo-Hermiticity and establishing the metric operator $\rho = e^{-\sum_i q_i S_i^z}$ that defines the physical inner product.
Numerical diagonalization of the conserved charges revealed the hallmark spectral signature of $\mathcal{PT}$ symmetry: eigenvalues are either entirely real or form complex conjugate pairs. Notably, we observed partial $\mathcal{PT}$ symmetry breaking, where low-energy states remain in the unbroken phase while higher-energy states transition to the broken phase—a feature rooted in the dominance of Hermitian longitudinal interactions in the ground-state manifold.
Finally, we provided an exact analytical expression for spin dynamics in both the unbroken and broken phases. In the former, spins exhibit coherent oscillations; in the latter, dynamics acquire exponential envelopes reflecting effective gain and loss. All observables were computed using the physical $\mathcal{CPT}$ inner product, ensuring norm conservation and real expectation values.

\begin{acknowledgments}
We thank the anonymous referees for their insightful comments and constructive suggestions, which significantly improved the clarity and rigor of this manuscript.
\end{acknowledgments}

\appendix
\section{Derivation of Integrable Couplings and Magnetic Fields}
\label{app:derivation}
We derive the explicit forms of the complex magnetic field components \(B_i^\alpha\) and anisotropic couplings \(\hat{\Gamma}_{ij}^\alpha\) (\(\alpha = x,y,z\)) that ensure mutual commutativity of the conserved charges:
\begin{equation}
    [Q_i, Q_j] = 0 \quad \forall\, i \neq j,
\end{equation}
where
\begin{equation}
    Q_i = \sum_{\alpha} B_i^\alpha S_i^\alpha + \sum_{k \neq i} \sum_{\alpha} \hat{\Gamma}_{ik}^\alpha S_i^\alpha S_k^\alpha.
\end{equation}

\subsection{Commutator Structure}

Using the spin algebra \([S_i^\alpha, S_j^\beta] = i \delta_{ij} \varepsilon^{\alpha\beta\gamma} S_i^\gamma\), the commutator \([Q_i, Q_j]\) contains three types of terms:
\begin{enumerate}
    \item \textbf{Two-site terms} (\(i,j\)): proportional to \(S_i^\alpha S_j^\beta\)
    \item \textbf{Three-site terms} (\(i,j,k\)): proportional to \(S_i^\alpha S_j^\beta S_k^\gamma\)
    \item \textbf{Single-site terms}: proportional to \(S_i^\alpha\)
\end{enumerate}
For integrability, all coefficients must vanish identically.

\subsection{Three-Site Terms and Anisotropy Constraints}

The three-site terms arise from products like \(\hat{\Gamma}_{ik}^\alpha \hat{\Gamma}_{jk}^\beta S_i^\alpha S_j^\beta S_k^\gamma\). Their cancellation requires the anisotropy condition \cite{Dimo,Faribault}:
\begin{equation}
    \hat{\Gamma}_{ik}^x \hat{\Gamma}_{jk}^y - \hat{\Gamma}_{ij}^x \hat{\Gamma}_{jk}^z - \hat{\Gamma}_{ji}^y \hat{\Gamma}_{ik}^z = 0,
\end{equation}
and cyclic permutations. This is satisfied if the couplings factorize as:
\begin{equation}
    \hat{\Gamma}_{ij}^\alpha = g \frac{f_\alpha(\epsilon_i) f_\beta(\epsilon_j)}{\epsilon_i - \epsilon_j}, \quad (\alpha,\beta,\gamma) \text{ cyclic},
\end{equation}
where \(\{\epsilon_i\}\) are distinct inhomogeneity parameters, and \(g\) is a global coupling constant.

\subsection{Two-Site Terms and Magnetic Field Constraints}

The two-site terms yield the condition:
\begin{equation}
    \hat{\Gamma}_{ij}^\beta B_j^\alpha + \hat{\Gamma}_{ji}^\alpha B_i^\alpha = 0, \quad \alpha \neq \beta.
\end{equation}
Assuming the factorized form above, this implies:
\begin{equation}
    B_i^\alpha = \frac{c_\alpha}{f_\alpha(\epsilon_i)},
\end{equation}
for constants \(c_\alpha\).

\subsection{Choice of Parametrization}

Following \cite{Dimo}, we choose square-root parametrizations to match the XYZ structure:
\begin{align}
    f_x(\epsilon) &= \sqrt{\alpha_x \epsilon + \beta_x}, \\
    f_y(\epsilon) &= \sqrt{\alpha_y \epsilon + \beta_y}, \\
    f_z(\epsilon) &= 1,
\end{align}
which ensures the correct anisotropy ratios. For the non-Hermitian case, we require \(B_i^x, B_i^y, \hat{\Gamma}_{ij}^x, \hat{\Gamma}_{ij}^y \in i\mathbb{R}\) (purely imaginary) to enable \(\mathcal{PT}\)-symmetry, while \(B_i^z, \hat{\Gamma}_{ij}^z \in \mathbb{R}\).

Thus, we set:
\begin{align}
    c_x &= i\delta, \quad c_y = i\lambda, \quad c_z = 1,
\end{align}
with \(\delta, \lambda, g \in \mathbb{R}\). This yields the final forms:

\subsection{Final Expressions}

The magnetic field components are:
\begin{align}
    B_i^x &= \frac{i\delta}{\sqrt{\alpha_x \epsilon_i + \beta_x}}, \\
    B_i^y &= \frac{i\lambda}{\sqrt{\alpha_y \epsilon_i + \beta_y}}, \\
    B_i^z &= 1.
\end{align}

The anisotropic coupling constants are:
\begin{align}
    \hat{\Gamma}_{ij}^x &= i g \frac{\sqrt{(\alpha_x \epsilon_i + \beta_x)(\alpha_y \epsilon_j + \beta_y)}}{\epsilon_i - \epsilon_j}, \\
    \hat{\Gamma}_{ij}^y &= i g \frac{\sqrt{(\alpha_y \epsilon_i + \beta_y)(\alpha_x \epsilon_j + \beta_x)}}{\epsilon_i - \epsilon_j}, \\
    \hat{\Gamma}_{ij}^z &= g \frac{\sqrt{(\alpha_x \epsilon_i + \beta_x)(\alpha_y \epsilon_i + \beta_y)}}{\epsilon_i - \epsilon_j}.
\end{align}

These expressions satisfy all integrability conditions \eqref{integra1}--\eqref{integra2} and reduce to the Hermitian XYZ Gaudin model when \(\delta = \lambda = 0\). The purely imaginary transverse fields and couplings enable \(\mathcal{PT}\)-symmetry, while the real longitudinal components preserve the physical structure of the model.

\end{document}